\documentclass[aps,pre,twocolumn,showpacs]{revtex4-1}
\usepackage{amsmath,graphicx,amsbsy,amssymb,amsmath,epsfig,latexsym,wasysym,pifont,mathrsfs} 
\usepackage{float,color} 
\newfloat{widefig}{thp}{lop}
\usepackage{comment} 
\usepackage{verbatim}
\usepackage{multirow,array}

\begin{document}
\title{Crossover from rotational to stochastic sandpile universality
  in the random rotational sandpile model} 
\author{Himangsu Bhaumik} \author{Jahir Abbas Ahmed} \author{S. B. Santra}
\email{santra@iitg.ac.in} \affiliation{Department of Physics, Indian
  Institute of Technology Guwahati, Guwahati-781039, Assam, India.}
\date{\today}

\begin{abstract}
  In the rotational sandpile model, either the clockwise or the
  anti-clockwise toppling rule is assigned to all the lattice
  sites. It has all the features of a stochastic sandpile model but
  belongs to a different universality class than the Manna class. A
  crossover from rotational to Manna universality class is studied by
  constructing a random rotational sandpile model and assigning
  randomly clockwise and anti-clockwise rotational toppling rules to
  the lattice sites. The steady state and the respective critical
  behaviour of the present model are found to have a strong and
  continuous dependence on the fraction of the lattice sites having
  the anti-clockwise (or clockwise) rotational toppling rule. As the
  anti-clockwise and clockwise toppling rules exist in equal
  proportions, it is found that the model reproduces critical
  behaviour of the Manna model. It is then further evidence of the
  existence of the Manna class, in contradiction with some recent
  observations of the non-existence of the Manna class.
\end{abstract}
\maketitle

\section{Introduction} 
A sandpile is a prototypical model to study self-organized criticality
(SOC) \cite{bak,*jensen,*pruessner}, which refers to the intrinsic
tendency of a wide class of slowly driven systems to evolve
spontaneously to a non equilibrium steady state characterized by
long-range spatiotemporal correlation and power-law scaling
behaviour. Several crossover phenomena from one sandpile universality
class to the other are reported in the literature on sandpile
models. For example, a crossover from Bak, Tang and Wiesenfeld (BTW)
\cite{btwPRL87} to the stochastic Zhang model was observed by O. Biham
     {\em et al.} \cite{bihamPRE01} by controlling the fraction of
     energy distributed to the nearest neighbours in a toppling. A
     crossover from the deterministic Zhang model \cite{zhangPRL89} to
     the stochastic sandpile model (SSM)
     \cite{mannaJPA91,mannaPHYA91,dharPHYA99b} was studied by L{\"
       u}beck \cite{lubeckPRE00b} by controlling the threshold
     condition. A crossover from the directed sandpile model (DSM)
     \cite{dharPRL89} to the directed percolation (DP) class
     \cite{dickman,*henkel,*DPhinrichsenAIP00} was observed by
     Tadi{\'c} and Dhar by introducing a stickiness parameter in the
     DSM \cite{tadicPRL97}. The crossover phenomena studied in these
     models are usually from a deterministic to a stochastic
     model. However, the universality class of a sandpile model is
     believed to be determined by the underlying symmetry present in
     the model \cite{rossiPRL00}. A crossover from one sandpile
     universality class to another then requires a change in the
     underlying symmetry of a given model. It is therefore intriguing
     to study a crossover phenomenon within the stochastic class of
     models with different symmetries in the toppling rule, and to
     look for spontaneous symmetry breaking in the system as one of
     the system parameters is tuned. Two such stochastic sandpile
     models are the SSM \cite{dharPHYA99b} and the rotational sandpile
     model (RSM) \cite{santraPRE07,ahmedPRE12}. The SSM is governed by
     externally imposed stochastic toppling rules. On the other hand,
     the RSM is governed by deterministic rotational toppling rules
     (except the very first toppling) and has broken mirror
     symmetry. Such a model can be useful in studying the avalanche
     dynamics of charged particles in the presence of a uniform
     magnetic field. In the RSM, an internal stochasticity appears due
     to a superposition of toppling waves from different directions
     during time evolution. Eventually, that induces all the features
     of a stochastic sandpile model, such as toppling imbalance,
     negative time auto correlation, and existence of finite-size
     scaling (FSS) into the RSM
     \cite{santraPRE07,ahmedEPJB10,ahmedCPC11,ahmedPRE12}. The RSM is
     thus a stochastic model, but it belongs to a completely different
     universality class than the Manna class of the SSM. The question
     is whether it would be possible to reproduce the critical
     behaviour of the SSM of Manna type in a model such as the RSM,
     which is stochastic due to its internal dynamics. Moreover, there
     is a long standing debate in the study of SOC as absorbing state
     phase transitions (APT)
     \cite{dickmanPRE98,munozPRE99,vespignaniPRE00} stating that the
     stochastic universality class or the Manna class is essentially
     the DP universality class. There is continuing evidences in
     favour
     \cite{bonachelaPRE06,*bonachelaPHYA07,*bonachelaPRE08,*leePRL13,*leePRE13,*leePRE14a,*leePRE14b}
     and against \cite{mohantyPRL02,*mohantyPHYA07,*basuPRL12} the
     existence of the so-called Manna class. If due to any external
     condition on the RSM, it reproduces the critical behaviour of the
     Manna model, that will be additional independent evidence for the
     existence of the Manna class.

In this paper, the crossover from one stochastic universality class to
another is studied by constructing a random rotational sandpile model
(RRSM) and the existence of the Manna class is discussed in the
context of random mixing of two conflicting rotational toppling rules
in the model.

\section{The model}
The RRSM is defined on a two-dimensional ($2D$) square lattice of size
$L\times L$. Initially, all lattice sites are assigned with the
clockwise toppling rule (CTR). A fraction $p$ of lattice sites are
then changed to the anti-clockwise toppling rule (ATR) randomly. The
toppling rules assigned to the lattice sites remain unchanged during
the time evolution of the system and hence this can be considered as a
quenched random configuration of the toppling rules. The RSM
\cite{santraPRE07} was defined for the presence of only one type of
toppling rule either CTR or ATR. Since the sandpile dynamics is
independent of the sense of rotations, the RSMs with either CTR or ATR
have the same critical behaviour.
\begin{figure}[t]
\centerline{ 
  \hfill \psfig{file=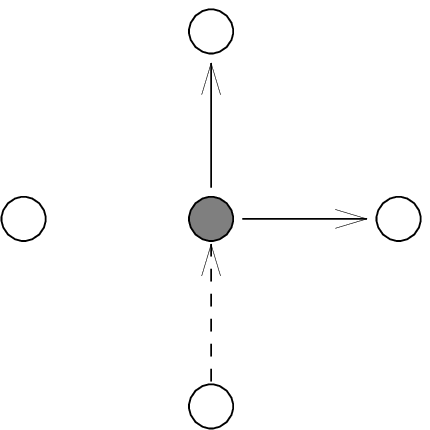,width=0.12\textwidth}
  \hfill \hfill \psfig{file=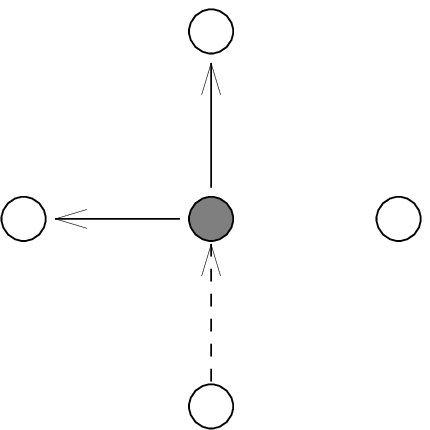,width=0.12\textwidth} 
  \hfill
   } 
\centerline{\hfill (a) CTR \hfill \hfill  (b) ATR \hfill}
\caption{\label{demo} The two toppling rules of RRSM, CTR, and ATR are
  demonstrated on a square lattice in (a) and (b), respectively. The
  active site in grey is at the center and it has received the last
  sand grain from the bottom (say), represented by a dashed arrow. The
  directions of sand flow from the active site are represented by
  solid arrows. In (a), due to CTR, one sand grain goes to the right
  and the other goes up, whereas in (b), due to ATR, one sand grain
  goes to the left and the other goes up. }
\end{figure} 

Each lattice site $i$, irrespective of the type of toppling rule it
has, is assigned with a positive integer $h_i$ representing the height
(the number of sand grains) of the sand column. Initially, all $h_i$s
are set to zero. The system is driven by adding sand grains, one at a
time, to randomly chosen lattice sites $i$. The critical height of the
model is taken as $h_c=2$. As the height of a sand column $h_i$
becomes equal to or greater than the critical height $h_c$, {\em
  i.e.}, $h_i\ge h_c$, the site becomes active and bursts into a
toppling activity. On the very first toppling of an active site, two
sand grains are given away to two randomly selected nearest neighbours
out of the four nearest neighbours on a square lattice. As soon as a
site $j$ receives a sand grain, the direction $d_j$ from which the
grain was received is assigned to it besides incrementing the height
of the sand column $h_j$ by one unit. The value of $d_j$ can change
from $1$ to $4$, as there are four possible directions on a square
lattice. The directions from an active site $i$ are defined as $d_i=1$
for left, $d_i=2$ for up, $d_i=3$ for right and $d_i=4$ for down. As
the avalanche propagates, the direction $d_i$ and height $h_i$ are
updated upon receiving a sand grain, and only the information
regarding the direction from which the last sand grain was received is
kept. The next active sites with $h_i\ge h_c$ in the avalanche will
topple following a deterministic rotational toppling rule. The
toppling rules for an active site $i$ that has received the last sand
grain from a direction $d_i$ are given below. If the active site $i$
is with CTR, the sand distribution is given by
\begin{equation}
\label{ctr}
{\rm CTR:} \ \ h_i \rightarrow h_i -2, \ \  h_j \rightarrow h_j+1,
\ \ \  j=d_i,d_{i+1},
 \end{equation} 
where one sand grain goes along $d_i$ and the other goes in a
clockwise direction with respect to $d_i$. If the active site $i$ is
with ATR, the sand distribution is given by
\begin{equation}
\label{actr}
{\rm ATR:} \ \  h_i \rightarrow h_i -2, \ \ h_j \rightarrow h_j+1,
\ \ \  j=d_i,d_{i-1},
 \end{equation} 
where one sand grain goes along $d_i$ and the other goes in an
anti-clockwise direction with respect to $d_i$. If the index $j$
becomes $5$, it is taken to be $1$; if it becomes $0$, it is taken to
be $4$. The CTR and ATR are demonstrated in Fig. \ref{demo} on a
square lattice. The avalanche stops if there is no active site present
and the system becomes under critical. The next sand grain is then
added. As RSM, the RRSM is non-Abelian \cite{dharPHYA99a,*dharPHYA06}
and it has no toppling balance \cite{karmakarPRL05}.

In the following, the results of the RRSM are compared with those of
the original RSM \cite{santraPRE07} and the SSM
\cite{dharPHYA99b}. The SSM considered here is a modified version of
the Manna model \cite{mannaJPA91,mannaPHYA91} known as the Dhar
Abelian model. The toppling rule in this SSM is that two sand grains
of an active site are given to two randomly selected nearest-neighbour
sites out of four possible nearest neighbours on a square lattice and
the height of the active site is reduced by $2$. The remaining sand
grains remain at the present site.

\begin{figure}[t]
\centering
\psfig{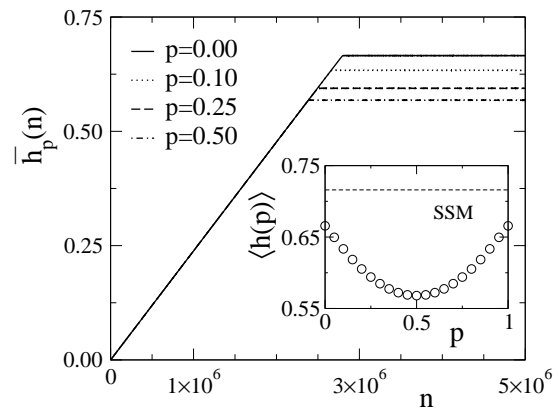}
\caption{\label{avght} Plot of ${\bar h}_p(n)$ against $n$, the number
  of sand grains added, for $p=0.0, 0.1, 0.25$ and $0.5$ for RRSM on
  $L=2048$. In the inset, the average steady state height
  $\langle h(p) \rangle$ is plotted against $p$. The
  dashed line represents the average steady state height of the SSM
  studied on $L=2048$.}
\end{figure}

\section{Steady state}
The steady state of a sandpile model corresponds to constant currents
of sand influx and outflux. Consequently, the average height of the
sand columns remains constant over time. For a given value of $p$ in
the RRSM, the mean height ${\bar h}_p(n)$ of the sand columns is
expected to be constant over the number of sand grains $n$ added
(equivalent to time) to the system. It can be defined as
\begin{equation}
\label{ah}
{\bar h}_p(n)   =\frac{1}{L^2}\sum_{i=1}^{L^2} h_i(p,n)
\end{equation} 
for a system size $L$. For different values of $p$, ${\bar h}_p(n)$
for $L=2048$ is plotted against $n$, the number of sand grains added
in Fig. \ref{avght}. After the addition of a sufficiently large number
of sand grains, the system reaches a steady state corresponding to a
given value of $p$. In order to study the effect of $p$ on the steady
state height, the saturated average height $\langle h(p) \rangle$ of
the sand columns in the steady state for a given value of $p$ is
estimated taking the average over the last $10^5$ sand grains on every
$64$ random configurations of toppling rules. Note that, no
configurational average is required for $p=0$ and $1$. In the inset of
Fig. \ref{avght}, $\langle h(p) \rangle$ is plotted against $p$ for
$L=2048$. It can be seen that the saturated average height of the sand
columns at the steady state decreases as $p$ is varied from $0$ (or
$1$) to $0.5$, and it attains a minimum value at $p=0.5$. The values
of $\langle h(p) \rangle$ are found to be symmetric about $p=0.5$, as
expected. The average heights at $p=0$ and $1$ are found to be that of
the RSM \cite{santraPRE07}, whereas at $p=0.5$ the average height is
that of the SSM. The time-averaged steady state-height for the SSM for
$L=2048$ is measured independently, and it is found to be $0.7162 \pm
0.0001$. It is represented by a dashed line in the inset of
Fig. \ref{avght}. It can be noted here that the measured value of the
average sand column height for the SSM is in good agreement with that
of the driven dissipative sandpile in the context of the precursor to
a catastrophe study \cite{pradhanPRE01} as well as the critical point
of the APT of a fixed energy sandpile model
\cite{vespignaniPRE00}. Thus, the steady-state heights corresponding
to different values of $p$ are not only different from each other but
also very different from that of the SSM.

\section{Results and discussion}
The critical properties of RRSM are characterized studying the
properties associated with avalanches in the steady states at
different values of $p$ and the system size $L$ on $2D$ square
lattices. The value of $p$ is varied from $0$ to $1$, and the system
size $L$ is varied from $128$ to $2048$ in multiples of $2$ for every
value of $p$. For the sake of comparison, data for the SSM are also
generated for the same lattice sizes. The information of an avalanche
is kept by storing the number of toppling of all the lattice sites in
an array $S_{L,p}[i],i=1,\cdots,L^2$ for given $L$ and $p$. All
avalanche properties of interest, such as the two point toppling
height correlation function, the toppling surface width, avalanche
size, etc., will be derived from $S_{L,p}[i]$.
\begin{figure}[t]
\centerline{
\psfig{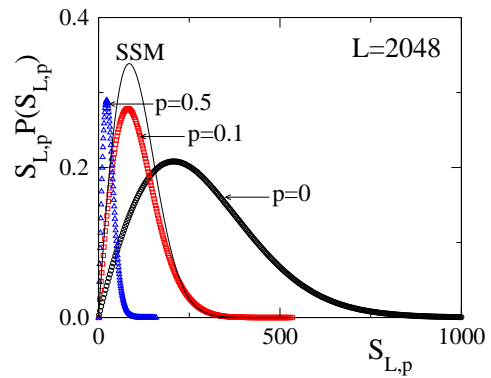}}
\caption{\label{sps}(Colour online) Plot of $S_{L,p}P(S_{L,p})$ versus
  $S_{L,p}$ for RRSM at different values of $p$ $(0.0, 0.1, 0.5)$ for
  $L=2048$. For the sake of comparison, the same distribution for the
  SSM is shown by a solid line. }
\end{figure}

\begin{figure}[t] 
  \centerline{\hfill
    \psfig{file=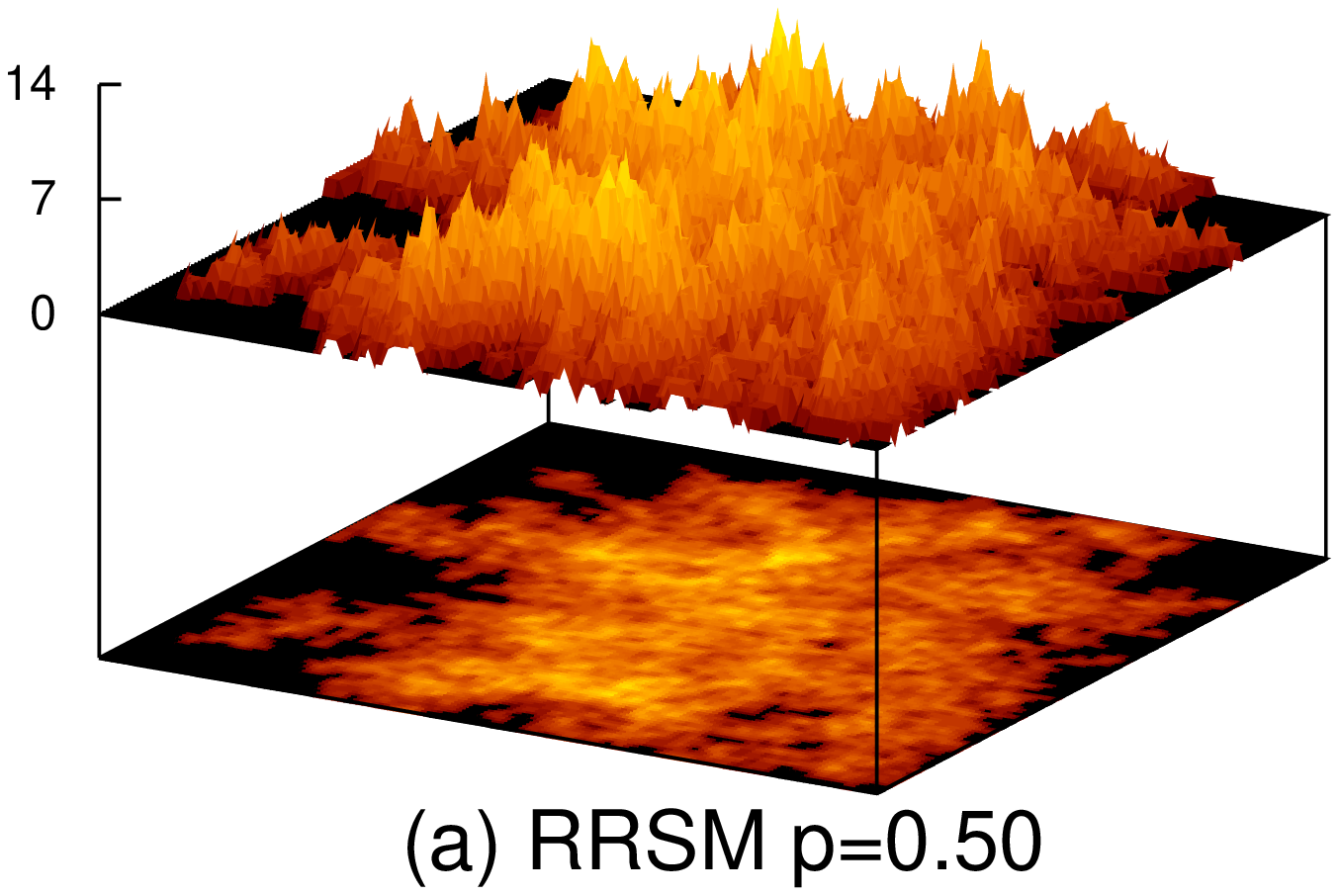,width=0.24\textwidth}
   \hfill \psfig{file=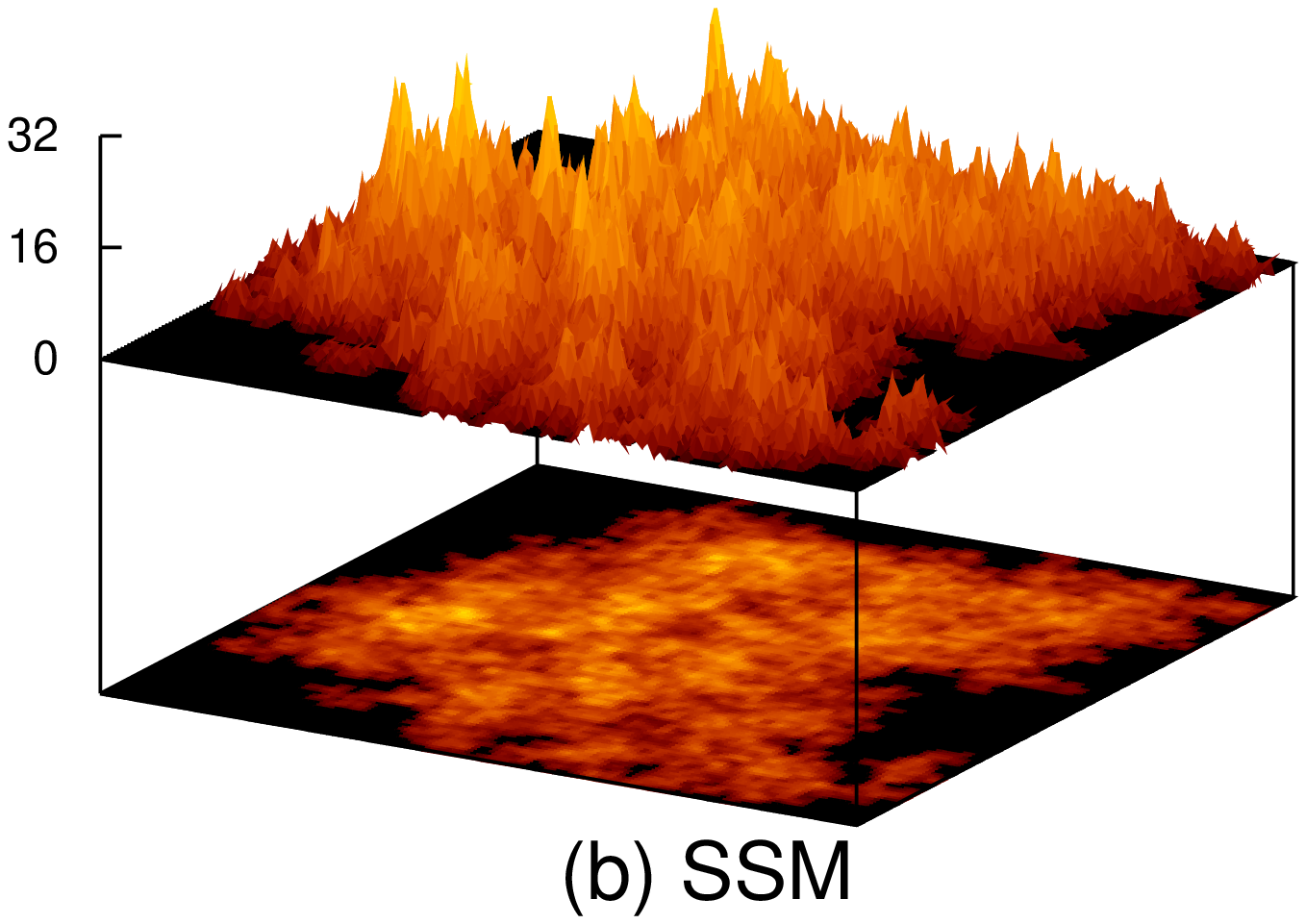,width=0.24\textwidth}
   \hfill
  } 
\caption{\label{tsurf}(Colour online) (a) Toppling surfaces of RRSM at
  $p=0.5$ generated on a square lattice of size $L=128$. (b) Toppling
  surface of the SSM generated on the same system size. The highest
  toppling number corresponds to a light brown colour and the lower
  toppling numbers are represented by darker and darker brown
  colours. Below each surface, their $2D$ projections are shown. These
  projections represent the avalanche clusters in two dimensions.}
\end{figure}

\subsection{Distribution of $S_{L,p}$ and avalanche morphology}
The probability distribution of $S_{L,p}$ is defined as
$P(S_{L,p})=n_S/A$ where $n_S$ is the number of sand columns that
toppled $S$ times and $A$ is the number of distinct sand columns (or
lattice sites) toppled. A FSS study of $P(S_{L,p})$ reported in
\cite{ahmedEPL10} suggests that RSM and SSM follow FSS, but BTW does
not. In Fig. \ref{sps}, distributions $S_{L,p}P(S_{L,p})$ of RRSM are
plotted against $S_{L,p}$ for several values of $p$ for a fixed
$L=2048$ and compared with that of the SSM. The distribution
$S_{L,p}P(S_{L,p})$ is found to be of Poisson type as
expected. However its height, width, and the peak position vary
strongly with $p$ on a given lattice. It can also be noted that the
distribution of the SSM is not identical with that of RRSM at
$p=0.5$. This implies that the internal structure of an avalanche at
different values of $p$ is different, but also it is different from
that of the SSM in comparison to that of RRSM at $p=0.5$. The
avalanche morphologies of two typical avalanches generated on a
$128\times 128$ square lattice for RRSM at $0.5$ and for that of the
SSM are presented in Fig. \ref{tsurf}. The values of the toppling
number at different lattice sites of an avalanche define a surface in
three dimensions called toppling surface \cite{ahmedEPL10}. Thus, the
height of the toppling surface at the $i$th lattice site is then given
by $S_{L,p}[i]$. The toppling surface of RRSM at $p=0.5$ is found to
be fluctuating all over the lattice as that of the SSM with different
maximum heights. For SSM, it is approximately $31$, whereas that for
RRSM at $p=0.5$ is approximately $12$, similar to the observation of
their steady-state heights. The projection of the toppling surfaces in
two dimensions is shown below the respective toppling surfaces. The
$2D$ view of a toppling surface is known as an avalanche cluster. It
can be seen that the avalanche cluster of RRSM at $p=0.5$ exhibits
random mixing of colours representing different toppling numbers as
that of an avalanche cluster of the SSM.  Note that both are very
different from that of the RSM in which a random superposition of
several BTW-type concentric zones \cite{bihamPRE01,grassbergerJPF90}
of lower and lower toppling numbers around different maximal toppling
zones is observed \cite{santraPRE07}. Since the avalanche morphologies
of RRSM at $p=0.5$ and the SSM are found to be similar, it is expected
that both models have the same critical behaviour, though the
distributions $S_{L,p}P(S_{L,p})$ of them are different.

\subsection{Properties of avalanche size}
One of the macroscopic avalanche properties is the total number of
toppling in an avalanche, called the toppling size $s$. Knowing the
values of $S_{L,p}[i]$ at every lattice site, the toppling size $s$ of
an avalanche can be obtained as
\begin{equation}
\label{ts}
s(L,p)=\sum_{i=1}^{L^2}S_{L,p}[i]
\end{equation}
for given $L$ and $p$. At the steady state, $5\times 10^5$ avalanches
are generated on every quenched random configuration of the toppling
rules for a given value of $p$. For every value of $p$, $64$
configurations of quenched toppling rules are considered. Therefore,
ensemble averaging is performed over $N_{\rm tot}=32\times 10^6$
avalanches for given values of $L$ and $p$. The probability to have an
avalanche of toppling size $s$ is given by the
$P_{L,p}(s)=N_{s}(L,p)/N_{\rm tot}$ where $N_{s}(L,p)$ is the number
of avalanches of toppling size $s$ for given $L$ and $p$ out of total
number of avalanches $N_{\rm tot}$ generated. For RSM (corresponding
to $p=0$ and $1$ of RRSM), it is already known that the distribution
of $s$ follows a power law scaling with a well defined exponent
$\tau_s$ and obey FSS \cite{santraPRE07,ahmedEPJB10}. The probability
distributions $P_{L,p}(s)$ for the toppling size $s$ for several
values of $p$ (other than $0$ and $1$) for a fixed large system sizes
$L=2048$ and for several values of $L$ for a fixed value of $p$ show
that not only the cutoffs but also the slopes of the distributions
depend on $p$ for a fixed $L$ whereas for a fixed $p$ only the cutoffs
of the distributions depend on $L$ keeping the slopes unchanged. For a
given $p$, a probability distribution function for toppling size $s$
is then proposed as
\begin{equation}
\label{pdlp}
P_{L,p}(s) = s^{-\tau_s(p)}f_{p}\left[\frac{s}{L^{D_s(p)}}\right],
\end{equation}
where $f_{p}$ is the $p$ dependent scaling function and $D_s(p)$ is
the capacity dimension of the toppling size $s$ corresponding to given
$p$. To have estimates of the exponents $\tau_s(p)$ and $D_s(p)$
defined in Eq. (\ref{pdlp}), the concept of moment analysis
\cite{karmakarPRL05,lubeckPRE00a} for the avalanche size $s$ has been
employed. For a given $p$, the $q$th moment of $s$ as function of $L$
is obtained as
\begin{eqnarray}
\label{qmnts1}
\langle s^q(L,p) \rangle &=& \int_0^{\infty} s^{q}P_{L,p}(s)ds
 \sim  L^{\sigma_s(p,q)},
\end{eqnarray}
\begin{figure}[t]
\centerline{
\psfig{file=santra_fig_5a.eps, width=0.22\textwidth}  
\psfig{file=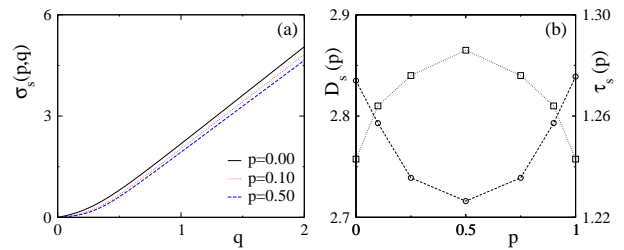, width=0.22\textwidth}}
\caption{\label{map}(Colour online) (a) Plot of $\sigma_s(p,q)$
  against $q$ for different values of $p$. (b) The variations of
  $\tau_s(p)$ and $D_s(p)$ against $p$ are shown. }
\end{figure}
where the moment scaling exponent $\sigma_s(p,q)$ would be
$\sigma_s(p,q)=D_s(p)[q+1-\tau_s(p)]$ for large $q$ as
$q>\tau_s(p)-1$. For each value of $p$, a sequence of values of
$\sigma_s(p,q)$ as a function of $q$ is determined by estimating the
slope of the plots of $\log\langle s^q(L,p)\rangle$ versus $\log(L)$
for $200$ equidistant values of $q$ between $0$ and
$2$. $\sigma_s(p,q)$ is plotted against $q$ for different values of
$p$ in Fig. \ref{map}(a). To obtain the values of $\tau_s(p$) and
$D_s(p)$, the direct method developed by L\"ubeck \cite{lubeckPRE00a} is
employed in which a straight line is fitted through the data points
for $q>\tau_s(p)-1$. From the straight line fitting, the $x$ intercept
provides $\tau_s(p)-1$ and the $y$-intercept provides
$D_s(p)[1-\tau_s(p)]$. Straight lines are fitted through the data
points for different values of $p$ in the range of $q$ between $1$ and
$2$, and the $x$ and $y$ intercepts are noted. The estimated values of
the exponents $\tau_s(p)$ and $D_s(p)$ are then presented in
Fig. \ref{map}(b) against $p$. There is a continuous change in the
values of $\tau_s(p)$ and $D_s(p)$ as $p$ changes from $0$ to
$1$. This indicates a continuous crossover of the critical behaviour
of the system through a series of universality classes of RRSM at
different values of $p$. The exponents $\tau_s(p)=1.234(13)$ and
$D_s(p)=2.82(2)$ at $p=0$ and $1$ correspond to that of the RSM
\cite{santraPRE07} as expected. However, the values of the critical
exponents $\tau_s(p)=1.286(9)$ and $D_s(p)=2.71(2)$ at $p=0.5$ are
found very close to that of the SSM
\cite{huynhJSM11,chessaPRE99,lubeckPRE00a}. Although both of the
exponents are varying continuously with $p$, due to the diffusive
behaviour of RRSM, the scaling relation $D_s(p)[2-\tau_s(p)]=2$ is
found to be valid within error bars for all values of $p$. It can be
noted here that the continuous crossover in RRSM from one stochastic
to another stochastic universality class through a series of
stochastic classes is very different from the observed crossover
phenomena from one deterministic to a stochastic universality class
\cite{karmakarPRL05,cernakPRE06}.

\begin{figure}[t]
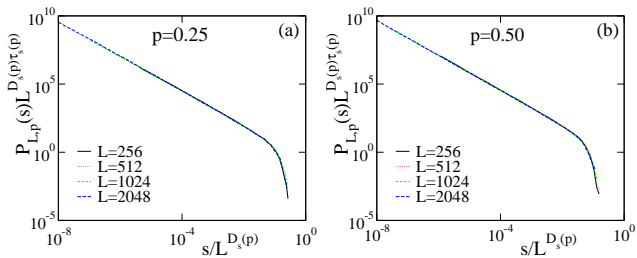

\centerline{
\psfig{file=santra_fig_6a.eps, width=0.23\textwidth} 
\psfig{file=santra_fig_6b.eps, width=0.23\textwidth}}
\caption{\label{mapc}(Colour online) Plot of
  $P_{L,p}(s)L^{D_s(p)\tau_s(p)}$ against $s/L^{D_s(p)}$ with
  different values of $L$ is shown in (a) for $p=0.25$ and in (b) for
  $p=0.50$. }
\end{figure}

To verify the form of the scaling function $f_p$ given in
Eq. (\ref{pdlp}), the scaled distributions
$P_{L,p}(s)L^{D_s(p)\tau_s(p)}$ is plotted against the scaled variable
$s/L^{D_s(p)}$ for four different system size $L$ for $p=0.25$ in
Fig.\ref{mapc}(a) and for $p=0.50$ in Fig. \ref{mapc}(b) in double
logarithmic scales. It can be seen for both the cases that data for
different values $L$ are collapsed onto a single curve, {\em i.e.},
the scaling function. Hence, the proposed scaling function form in
Eq. (\ref{pdlp}) for RRSM for any values of $p$ and $L$ is a correct
scaling function form. The analysis not only provides estimates of the
values of the exponents, but it also confirms that the model exhibits
FSS. The steady-state event distribution of this slowly driven
dynamical system is then found to obey power-law scaling at different
values of $p$; RRSM then exhibits SOC for any value of $p$ with
different sets of critical exponents. Moreover, at $p=0.5$ of RRSM,
the appearance of SSM confirms the existence of the Manna class.

\subsection{Properties of toppling surfaces}
\begin{figure}[t]
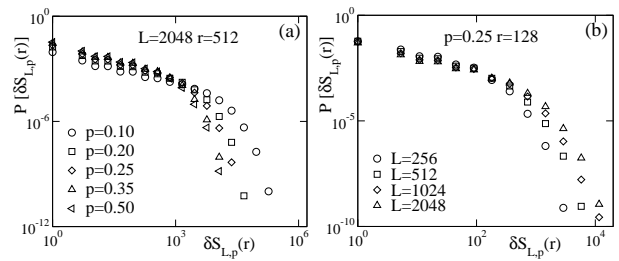

\centerline{
\psfig{file=santra_fig_7a.eps,width=0.22\textwidth}
\psfig{file=santra_fig_7b.eps,width=0.22\textwidth}}
\caption{\label{ppsr} (a) Probability distribution of $\delta
  S_{L,p}(r)$ for RRSM with $p=0.10$ ($\Circle$), $0.20$ ($\Box$),
  $0.25$ ($\Diamond$), $0.35$ ($\triangle$), $0.50$ ($\triangleleft$)
  with $L=2048$ and $r=512$. (b) Distribution of $\delta S_{L,p}(r)$
  for different $L$ as $L=256$ ($\Circle$), $512$ ($\Box$), $1024$
  ($\Diamond$), $2048$ ($\triangle$) for fixed values of $p=0.25$
  and $r=128$. }
\end{figure}

The toppling surfaces are obtained for large spanning avalanches
only. The spanning avalanches are those that are touching the opposite
sides of the given lattice. For a given system size $L$, a total of
$N_{\rm span}=1024$ spanning avalanches are taken over $64$ initial
random configurations for each values of $p$. To study the scaling
behaviour of a two-point height-height correlation function, the
correlation between the toppling numbers of two lattice sites
separated by a certain distance has to be determined. Since the
correlation function has to be calculated as a function of a
continuous variable $r$, (the distance between any two lattice sites),
the toppling number of a site is represented as $S_{L,p}(x)$, where $x$
is the position vector of the lattice site with respect to the origin
of a $2D$ coordinate system instead of using a discrete sequence of
toppling numbers stored in $S_{L,p}[i]$. The square of the difference
of toppling numbers $\delta S_{L,p}(r)$ at two lattice sites separated
by a distance $r$ is given by
\begin{equation}
\label{cr2}
\delta S_{L,p}(r)=\left|S_{L,p}(x + r)-S_{L,p}(x)\right|^2,
\end{equation}
where $S_{L,p}(x+r)$ is the toppling number at $x+r$ for given $L$ and
$p$. The probability $P[\delta S_{L,p}(r)]$ of a particular value of
$\delta S_{L,p}(r)$ occurring for a particular $r$ for a given value
of $L$, $p$ is defined as
\begin{equation}
\label{prds}
P[\delta S_{L,p}(r)] = \frac{n_{\delta S_{L,p}(r)}}{N_r},
\end{equation}
where $n_{\delta S_{L,p}(r)}$ is the number of pairs of sand columns
having the desired value of $\delta S_{L,p}(r)$ and $N_r$ is the total
number of pairs separated by a distance $r$ for all the surfaces. To
determine $N_r$, for each surface $100$ centers are randomly
selected. From each center, all possible sand columns at a distance
$r$ are counted and then added for $N_{\rm span}=1024$ surfaces. The
probability distribution $P[\delta S_{L,p}(r)]$ is then estimated for
several values of $L$, $p$, and $r$. To guess the form of the
distribution function $P[\delta S_{L,p}(r)]$, once is plotted against
$p$ for a fixed system size $L=2048$ and $r=512$ and then it is
plotted against $L$ for a fixed value of $p=0.25$ and $r=128$ in
Fig. \ref{ppsr}(a) and \ref{ppsr}(b), respectively. It can be seen
that as $p$ increases, the cutoffs of the distributions decrease for a
given system size $L$. On the other hand, for a given $p$ the cutoffs
increase as the system size $L$ is increased. Hence, following
Ref. \cite{alavaJSM06,*bouchbinderPRL06,*santucciPRE07,*bakkePRE07}
the form of the probability distribution function $P[\delta S(r,p,L)]$
is proposed as
\begin{equation}
\label{psrpL}
P[\delta S_{L,p}(r)]=\frac{r^{-2H(p)}}{L^\zeta}
g\left[\frac{\delta S_{L,p}(r)}{L^{\zeta}r^{2H(p)}}\right],
\end{equation} 
where $H(p)$ is the $p$-dependent Hurst exponent, $\zeta$ is another
exponents, and $g$ is the scaling function.

The correlation between the toppling numbers of two sand columns
separated by a distance $r$ can be obtained by estimating the
expectation $\langle \delta S_L(r,p) \rangle$. The correlation
function for a given $L$ and $p$, is obtained as
\begin{eqnarray}
\label{crp}
C_{L,p}(r) &=&\int_0^\infty \delta S_{L,p}(r)P[\delta S_{L,p}(r)]
d[\delta S_{L,p}(r)] \nonumber\\ 
&=& r^{2H(p)}L^{\zeta}\int_0^\infty zg(z)dz \nonumber\\
&\sim & r^{2H(p)}L^{\zeta}
\end{eqnarray}
where $z=\delta S_{L,p}(r)/(L^{\zeta}r^{2H(p)})$ is the scaled variable
and the value of the integral would be a constant. Notice that
$C_{L,p}(r)$ is a system size dependent correlation function. Such
correlation functions also appear in the cases of growing interfaces
in random media and self-affine fracture surfaces
\cite{lopezPRE95,*lopezPRE96,*lopezPRE98,*morelPRE98,*lopezPRL99}. In
order to determine the Hurst exponent $H(p)$, and the values of
$\zeta$, integrated correlation function $I_{L,p}(R)$ up to
a distance $R$ and the overall surface width $W_{L,p}$ are
estimated. $I_{L,p}(R)$ and $W_{L,p}$ are obtained as
\begin{eqnarray}
\label{sicrf}
I_{L,p}(R) &=& \int_0^R C_{L,p}(r)dr \sim 
R^{1+2H(p)}L^{\zeta}
\end{eqnarray}
and 
\begin{eqnarray}
\label{wc}
W^2_{L,p} &=& \frac{1}{L^2}\int_0^L C_{L,p}(r) rdr 
      \sim  L^{2\chi(p)},
\end{eqnarray}
\begin{figure}[t]
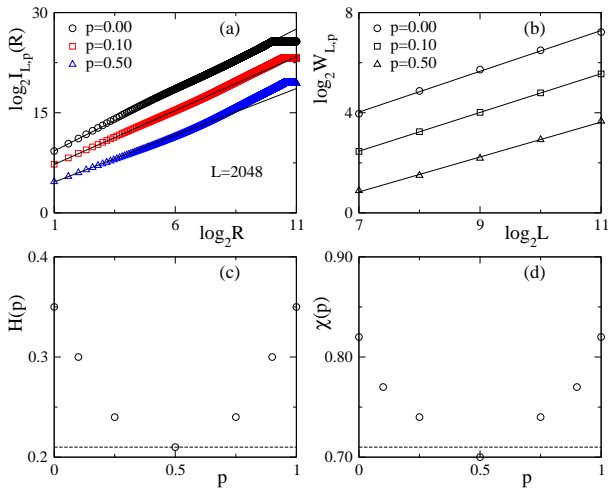

\centerline{
  \psfig{file=santra_fig_8a.eps,width=0.22\textwidth}
  \psfig{file=santra_fig_8b.eps,width=0.22\textwidth}}
\centerline{
  \psfig{file=santra_fig_8c.eps,width=0.22\textwidth}
  \psfig{file=santra_fig_8d.eps,width=0.22\textwidth}}
\caption{\label{ir}(Colour online) (a) Plot of $I_{L,p}(R)$ against
  $R$ for $p=0.0$ ($\Circle$), $p=0.10$ ($\Box$) and $p=0.50$
  ($\triangle$) on a system of size $L=2048$. The straight lines
  represent the linear least-squares-fitted lines through the linear
  region of the data points. (b) Plot of $W_{L,p}$ against $L$ for the
  same values of $p$ as in (a) with the same symbols. The straight
  lines are the linear least-squares-fitted lines through the data
  points. In (c) and (d), $H(p)$ and $\chi(p)$ are plotted,
  respectively, against $p$. The dashed lines in (c) and (d) represent
  the values of $H$ and $\chi$ for the SSM. }
\end{figure}
where $\chi(p)=\zeta/2+H(p)$ is known as the roughness exponent.
$I_{L,p}(R)$ against $R$ and $W_{L,p}$ against $L$ are plotted in
double logarithmic scales in Figs. \ref{ir}(a) and \ref{ir}(b),
respectively for the RRSM at $p=0, 0.1$ and $0.5$ and for different
values of $L=128-2048$. It can be seen that both $I_{L,p}(R)$ and
$W_{L,p}$ follow power-law scaling with their respective
arguments. The slopes are obtained by a linear least-squares fit
through the data points, and one obtains $1+2H(p)$ from (a) and
$\chi(p)$ from (b). The values of the Hurst exponents $H(p)$ and the
roughness exponent $\chi(p)$ are plotted against $p$ in
Figs. \ref{ir}(c) and \ref{ir}(d), respectively. A few observations
are there. First, a continuous crossover in the values of the critical
exponents has occurred as $p$ changed from $0$ (or $1$) to $0.5$. This
confirms the existence of a series of stochastic universality classes
as observed in the case of an avalanche size distribution
exponent. Second, not only the values of $H(p)\approx 0.35$ and
$\chi(p)\approx 0.82$ for the RRSM at $p=0$ and $1$ is found to be the
same as those of the RSM \cite{ahmedPRE12}, but also the values of
$H(p)\approx 0.21$ and $\chi(p) \approx 0.70$ for the RRSM at $p=0.5$
are found to be close to those of the SSM \cite{ahmedPRE12}. The
dashed lines in Fig. \ref{ir}(c) and \ref{ir}(d) represent the values
of $H$ and $\chi$ for the SSM. Therefore, the Manna class exists in
the strong disordered limit of the RRSM. Third, comparing the values
of $\chi(p)$ and $H(p)$ for all values $p$, it is observed that
$\chi(p) \approx H(p) + 1/2$, which suggests that $\zeta \approx
1$. In that case, the values of $\chi(p)$ and $H(p)$ obtained in the
RRSM for different values of $p$ do not satisfy the usual
Family-Vicsek scaling (no difference in the roughness exponent and the
Hurst exponent) \cite{familyJPA85}, rather they satisfy an anomalous
scaling given by $\chi(p)=\zeta/2+H(p)$ with $\zeta=1$ for any value
of $p$. Such a scaling relation is already verified for the RSM and
the SSM \cite{ahmedPRE12,ahmedPHYA12} independently. Note that the
anomalous scaling resulted here due to the system size dependence
of the correlation function $C_{L,p}(r)$ \cite{hansen}. Finally, the
critical exponent of toppling surfaces and that of the avalanche size
distribution are found to be related as $D_s(p)=2+\chi(p)$
\cite{ahmedPHYA12} at each value of $p$. The results obtained in two
different methods are then consistent.
\begin{figure}[t]
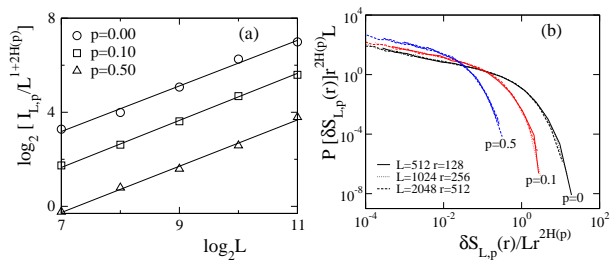

\centerline{\psfig{file=santra_fig_9a.eps,width=0.22\textwidth}
  \psfig{file=santra_fig_9b.eps,width=0.22\textwidth}}
\caption{\label{IRscaled}(Colour online) (a) Plot of
  $I_{L,p}/L^{1+2H(p)}$ against $L$ for different $p$. The straight
  line through data points gives the value of $\zeta \approx 1$. (b)
  Plot of $P[\delta S_{L,p}(r)]r^{2H(p)}L$ against $\delta
  S_{L,p}(r)/Lr^{2H(p)}$ for $p=0.0$, $p=0.1$, and $p=0.5$ for
  different values of $L$ and $r$ (see the legend).}
\end{figure}

To verify the scaling form of the probability distribution $P[\delta
  S_{L,p}(r)]$, the value of the exponent $\zeta$ must be
determined. To obtain the numerical value of $\zeta$,
$I_{L,p}/R^{1+2H(p)}$ for $R=L$ are plotted against $L$ for several
values of $p$ in Fig.\ref{IRscaled}(a). From the slopes of the linear
least-squares-fitted straight lines, it is found that $\zeta\approx 1$
for all values of $p$, as already predicted by the scaling
relations. The scaling function form given in Eq. (\ref{psrpL}) is now
verified by plotting a scaled distribution $P[\delta
  S_{L,p}(r)]r^{2H(p)}L$ against a scaled variable $\delta
S_{L,p}(r)/Lr^{2H(p)}$ for different values of $p$ in
Fig. \ref{IRscaled}(b). For $\zeta=1$, good data collapses are
observed for different values of $L$ and $r$ at three different values of
$p$ with the respective values of the Hurst exponent $H(p)$. Thus the
proposed scaling form assumed in Eq. (\ref{psrpL}) is correct. Such
distribution functions will then be useful to analyze rough surfaces
arising in a system with controlled disorder.

\subsection{Comparison of SSM and RRSM at $p=0.5$}
\begin{figure}[t]
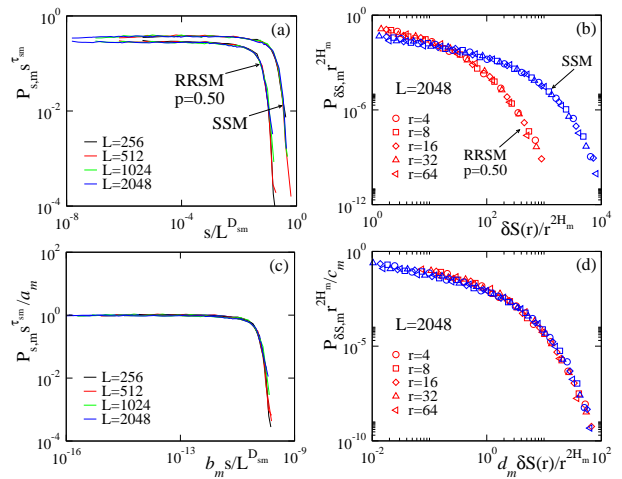

\centerline{ \psfig{file=santra_fig_10a.eps,width=0.22\textwidth}
\psfig{file=santra_fig_10b.eps,width=0.22\textwidth} }
\centerline{\psfig{file=santra_fig_10c.eps,width=0.22\textwidth} 
\psfig{file=santra_fig_10d.eps,width=0.22\textwidth}}
\caption{\label{compare_ps}(Colour online) (a) Plot of scaled toppling
  size distributions $P_{s,m}s^{\tau_{sm}}$ of the two models RRSM at
  $p=0.5$ and SSM against the scaled variable $s/L^{D_{sm}}$ for
  different system sizes $L=256$ (black line), $512$ (red line),
  $1024$ (green line), and $2048$ (blue line). (b) Plot of $P_{\delta
    S,m}r^{2H_m}$ against $\delta S(r)/r^{2H_m}$ for different values
  of $r$ (see the legend) for $p=0.5$ of the RRSM and the SSM for
  $L=2048$. The same distributions given in (a) and (b) are plotted in
  (c) and (d), respectively, after normalizing by the respective
  non-universal metric factors. }
\end{figure}

Alhough the macroscopic parameters describing the critical states of
the RRSM at $p=0.5$ and the SSM are found to be drastically different,
the values of the critical exponents are found to be similar. It is
then important to compare the scaling function forms of both
models. The probability distributions of toppling size $s$ and that of
the square difference of toppling numbers $\delta S(r)$ are considered
for comparison of their associated scaling functions. The
model-dependent probability distributions for $s$ and $\delta S(r)$
are proposed as
\begin{subequations}
\begin{equation}
\label{pdl_met}
P_{s,m} = a_{m}s^{-\tau_{sm}}f_{m}(b_{m}s/L^{D_{sm}})
\end{equation}
and 
\begin{equation}
\label{psr_met}
P_{\delta S,m}=c_mr^{-2H_m} g_m[d_m\delta S(r)/r^{2H_m}],
\end{equation}
\end{subequations}
where $m$ represents the models RRSM at $p=0.5$ and the SSM, and
$a_m$, $b_m$, $c_m$, and $d_m$ are non-universal metric factors that
contain all non-universal model-dependent features such as the lattice
structure, the update scheme, the type or range of interaction, etc.
\cite{lubeckPRL03,*lubeckPRE03}. The $p$ dependence in both the
functions and the $L$ dependence in Eq. (\ref{psr_met}) are dropped
because they are considered either for a fixed value of $p$ or for a
fixed value of $L$. Assuming $a_m$, $b_m$ and $c_m$, $d_m$ all are
equal to $1$, the scaled distributions $P_{s,m}s^{\tau_{sm}}$ for
several values of $L$ and $P_{\delta S,m}r^{2H_m}$ for several value
of $r$ for $L=2048$ are plotted in Figs. \ref{compare_ps}(a) and
\ref{compare_ps}(b) respectively, against their respective scaled
variables $s/L^{D_{sm}}$ and $\delta S(r)/r^{2H_m}$ using the same
values of $\tau_{sm}$, $H_m$ and $D_{sm}$ for both models. It can be
seen that in both cases, the distributions collapse onto two different
curves corresponding to the RRSM at $p=0.5$ and the SSM. It seems that
scaling functions are different for these two models, although they
scale independently with their respective arguments with the same
critical exponents. It is then essential to verify whether the scaling
functions are affected by the non-universal metric factors or not. The
values of $a_{m}$ are calculated from the limiting value of
$P_{s,mL}s^{\tau_{sm}}$ as $s\rightarrow 0$, and they are found to be
$a_{m}=0.34\pm0.01$ for the RRSM at $p=0.5$ and $a_{m}=0.27\pm0.01$
for the SSM. The values of $b_{m}$ are calculated from the average
toppling numbers $s$ of the two models for a fixed $L$. For $L=2048$,
the values are $b_{m}=2^{-16.92\pm0.02}$ for the RRSM at $p=0.5$ and
$b_{m}=2^{-18.65\pm0.03}$ for the SSM. The error bars represent the
uncertainty due to different independent runs. Similarly, the values
of $c_m$ are calculated from the limiting value of $P_{\delta S,m}$ as
$\delta S(r)\rightarrow 0$ for $r=1$. For the RRSM at $p=0.5$ and for
the SSM, the estimated values are $c_m=0.60\pm0.01$ and $0.25\pm0.01$,
respectively. The values of $d_m$ are essentially the inverse of the
averages of the scaled variable $\langle \delta S(r)/r^{2H_m} \rangle$
for the respective models. The values of $d_m$ are $2^{-3.66\pm0.02}$
and $2^{-6.93\pm0.02}$ for the RRSM at $p=0.5$ and the SSM,
respectively. In Fig. \ref{compare_ps}(c) and \ref{compare_ps}(d), the
scaled distribution functions are plotted against their scaled
variables incorporating the metric factors of the respective models
for both distributions. It can be seen that a reasonable data collapse
is observed in both cases. Hence, both the scaling functions $f_{sm}$
and $g_m$ for two models are essentially the same apart from the
non-universal metric factors associated with them. Therefore, the
critical state of the RRSM at $p=0.5$ belongs to the so called
stochastic or Manna universality class, although the origins of such a
universality class in the two models are completely different. The
stochasticity in the RRSM at $p=0.5$ is due to the simultaneous
presence of two conflicting toppling rules (CTR and ATR) randomly at
equal proportions, whereas the stochasticity in the SSM is externally
imposed through the toppling rules. This is therefore independent
confirmation of the existence of the Manna universality class as it is
observed in other models \cite{leePRL13,leePRE13} in the context of
APT on a diluted lattice.

\section{Conclusion}
A continuous crossover from RSM to SSM (Manna class) is observed in a
random rotational sandpile model as the fraction $p$ of lattice sites
with the anti-clockwise toppling rule (and the rest of the lattice
sites are with the clockwise toppling rule) varies from $0$ (or $1$)
to $0.5$. As $p$ changes from $0$ (or $1$) to $0.5$, the system passes
through a series of non-universal stochastic models at each value of
$p$.  Finally at $p=0.5$, at which there is maximum disorder in the
toppling rule, the RRSM corresponds to the Manna class. However, not
only does the origin of stochasticity in the RRSM and SSM differ, but
the macroscopic parameters identifying the critical steady states of
these models differ significantly as well. A scaling theory for such a
continuous crossover is developed and verified numerically by
estimating a set of critical exponents related to the avalanche
properties as well as to that of the toppling surfaces. The values of
the critical exponents satisfy all scaling relations among them for
all values of $p$. This study then not only represents a continuous
crossover from the RSM to the SSM, but it also confirms the existence
of the Manna class at the strong disorder limit of the RRSM.

\section*{Acknowledgments}
HB thanks the financial support from the Department of Science and
Technology, Ministry of Science and Technology, Government of India
through project No.SR/S2/CMP-61/2008.

\bibliographystyle{apsrev4-1}%apsrev4-1
\bibliography{ref}

\end{document}